\documentclass[manuscript]{aastex}

\begin{document}
\pagenumbering{arabic}

\title{Some Systematics of Galactic Globular Clusters}

\author{Sidney van den Bergh}
\affil{Dominion Astrophysical Observatory, National Research Council of Canada, 5071 West Saanich Rd.,
Victoria, B.C. V9E 2E7, Canada}

\begin{abstract}

The global properties of all known Galactic globular clusters are examined. The relationship between the luminosities and the metallicities of Galactic globular clusters is found to be complex. Among luminous clusters there is a correlation in the sense that the oldest clusters are slightly more metal deficient than are younger clusters. However, no such clear-cut relationship is found among the faintest globular clusters.
The central concentration index C of globular clusters is seen to be independent of metallicity. The dependence of the half-light radii of globular clusters on their Galactocentric distances can be approximated by the relation $R_h~ \alpha~ R^{2/3}_{gc}$. Clusters with collapsed cores are mostly situated close to the Galactic nucleus. For $R_{gc} < 10$ kpc the luminosities and the radii of clusters appear to be uncorrelated.
The Galaxy differs from the LMC and the SMC in that it appears to lack highly flattened luminous clusters. Galactic globular clusters with ages $\geq$ 13.0 Gyr are all of Oosterhoff type II, whereas almost all of those with ages $<$ 13.0 Gyr have been assigned to Oosterhoff type I. Globular clusters with ages $<$11.5 Gyr are all located in the outer Galactic halo, have below-average luminosities and above-average radii. On the other hand the very old globular cluster NGC 6522 is situated close to the Galactic nucleus.

\end{abstract}


\section{Introduction}

From a study of the distribution of 69 globular clusters Harlow Shapley (1918) concluded that these clusters were ``subordinate to the general [G]alactic system'' and that they cluster around what is now known to be the Galactic nuclear bulge. Later Helen Sawyer Hogg (1959) presented a catalog of
118 clusters that she considered to be probable globular clusters.
Subsequently Arp (1965) published a catalog of data on 119 globulars.
This was followed by a catalog of 129 globular clusters by Kukarkin (1974). Structural parameters for 154 globular clusters were listed by Webbink (1985) and a list of 143 probable globular clusters was published by Djorgovski \& Meylan (1993).
This catalog was subsequently superseded by the catalog of Harris (1996). Updates to the latter catalog were made available at http://physics.mcmaster.ca/resources/globular.html
in 2003, and most recently in December 2010. The fact that the number of recognized Galactic globulars only increased from 150 to 157 between 2003 and 2010 suggests that our sample of Galactic globular clusters is asymptotically approaching completeness. Data on, or derived from, Harris's (2010) catalog are listed in Table 1. It should, however, be noted that at least three of the objects contained in this table are probably not typical globular clusters, but rather the remnant cores of dwarf spheroidal galaxies. Both Omega Centauri (Bekki \& Freeman 2003) and Terzan 5 (Ferraro 2011) exhibit such a large range in iron abundances that they must have formed in a deep galactic potential well. However, Bailin \& Harris (2009) caution that the metal retention efficiency of massive young globulars will depend not only on their mass, but also on the strength of the tidal field in which they orbit, and on the density of the ultraviolet radiation field in which they are bathed. Furthermore Ibata, Gilmore \& Irwin (1994) have shown that M54 = NGC6715 is the core of the Sagittarius dwarf spheroidal galaxy. Other clusters that have been suspected of being the cores of (now mostly dissipated) dwarf spheroidal galaxies are NGC 1851 (Olszewski et al. (2009) and NGC5694 (Correnti et al.
2011). Recently Kaposov et al. (2007) have discovered two objects that appear to be extremely dim globular clusters which have such low masses that their evaporation time is only $\sim$ 0.1 times the age of the Galaxy. Many more such clusters might once have populated the Galactic halo.   Another recently discovered extremely faint globular cluster is SEGUE 3 that was recently discovered by Fadely et al. (2011).  Finally Minniti et al. (2011) have very recently discovered what might possibly be a faint new globular cluster (VVV CL001) in the Galactic bulge. These objects have not yet been incorporated into Table 1. For some previous work on the systematics of Galactic globular clusters the reader is referred to papers by Djorgovski (1995), Bellazzini (1998) and by Pasquato \& Bertin (2010).

Table 1 lists the following parameters for each Galactic globular cluster: The Galactocentric distance R measured in kpc, Log $R_{gc}$, the logarithm of the cluster metallicity relative to that of the Sun [Fe/H], the cluster absolute magnitude in visual light $M_v$, the cluster ellipticity (a-b)/a, where a and b are the cluster major and minor diameters, respectively, the concentration parameter C = Log $r_{t}/r_{c}$, where $r_t$ and $r_c$ are the the cluster tidal and core radii as defined by King (1966) and $R_h$ and Log $R_h$, where $R_h$ is the cluster half-light radius measured in pc. This parameter has been chosen because it is relatively insensitive to the effects of dynamical evolution (Spitzer \& Thuan 1972, Lightman \& Shapiro 1978, Murphy et al. 1990). Also given it Table 1 are cluster ages in Gyr by Dotter et al. (2010). Unfortunately such age information is available for fewer than half of the clusters in Table 1.
Finally the table lists the Oosterhoff type of each globular cluster for which this information is available.

\section{RELATIONS BETWEEN PARAMETERS}
\subsection{Galactocentric distance}

Figure 1 shows a plot of Galactocentric distance versus the
concentration parameter C (King 1966).   Clusters with collapsed cores do not fit King models.  For such clusters the concentration parameter was arbitrarily set to be C = 2.5.  This plot shows that (1) collapsed core clusters are strongly concentrated towards the Galactic center. For 21 such collapsed core clusters $<log R_{gc}>$ = 0.30, corresponding to $ R_{gc}$ = 1.98 kpc. [The very star-poor clusters Pal 1 ($M_v$ = -2.5) and Pal 12 ($M_v$ = -4.5) have $C > 2.5$. These high values are probably due to the difficulty of measuring the concentration index in such star-poor objects.] (2) Clusters with $R_{gc} >$ 30 kpc are significantly less compact than are those at smaller Galactocentric distances. This probably indicates that these outer halo clusters had a different evolutionary history from those that are associated with the main body of our Milky Way system.
Figure 2 shows that the half-light radii of Galactic globular clusters scatter widely around a relation (van den Bergh 1994) of the form

                  ~~~~~~~~~~~~~~~~~~~~~~~~ $R_h~ \alpha~  R^{2/3}_{gc}$~~~~~~~~~~~~~~~~~~~~~~~~~~~~~~~~~~~~~~~~~~~~~~~~~~~~~~~~.              (1)

This relationship has probably been imposed by the Galactic tidal field, which is determined by both $R_{gc}$ and the ratio of the cluster mass to the Galactic mass interior to $R_{gc}$, but also depends on the shape of the globular cluster's orbit (van den Bergh 1994, Correnti et al. 2011). Other factors contributing to the observed scatter are
(1) that the sizes of clusters are probably affected by dynamical interactions that are determined by their actual orbits, rather than by their present Galactocentric distances, (2) structural differences between normal and collapsed clusters and (3) observational 'noise'. Surdin (1994) has emphasized that the lack of large clusters at small values of $R_{gc}$ may, at least in part, be due to the destruction of fragile extended clusters by Galactic tidal forces. However, the absence of compact clusters the from the Galactic halo must mean that such objects never existed outside the main body of the Galaxy. The globular clusters in the Large Magellanic Cloud (van den Bergh 2000b) also exhibit a correlation between their sizes and their projected distances from the center of the LMC. However, the LMC clusters fall along the upper envelope of the radius versus Galactocentic distance relation observed for globular clusters associated with the Milky Way system. The globular clusters associated with the Fornax dwarf galaxy are all smaller than would be expected from their 138 kpc distance from the Galactic nucleus. This implies that their sizes were determined by distance from Fornax, rather than by their distance from the Galaxy. These results suggest (van den Bergh 2000b) that the Galaxy, the LMC and Fornax were already distinct stellar systems at the time that they started to form their globular clusters.

Not unexpectedly Figure 2 also confirms that the clusters with
collapsed cores are concentrated at smaller Galactocentric distances, than are those that have not yet collapsed. The most deviant cluster is the heavily crowded and reddened cluster HP 1 for which the half-light radius might have been overestimated. Alternatively (Ortolani et al. 1997) HP 1 could be a halo cluster that is presently passing through the Galactic bulge, or perhaps it might be an object that has been partly disrupted by tidal forces. Very recently Ortolani et al. (2011) have obtained adaptive optics images of HP1 from which they conclude that this objects may be the oldest globular cluster in the Galaxy and that it is spatially confined to the Galactic bulge. In fact, it might be the globular cluster that is situated closest to the Galactic nucleus.

Table 2 lists the clusters which, according to Dotter et al.
(2010), are younger than 11.5 Gyr. Inspection of this table shows that these young clusters are: (1) All located in the outer Galactic halo, (2) have above-average radii and (3) exhibit below-average luminosities.
Clearly these young globular clusters have had a very different evolutionary history from their older Galactic counterparts. The hypothesis that the clusters listed in Table 2 are unusual because they might originally have in formed in dwarf galaxies, that were subsequently captured by the Milky Way System, appears to be too simpleminded. Although The Sagittarius companion Terzan 7 (age 8.0 Gyr) is young its other probable comanions (Ibata et al. 1994) such as Arp 2 (13.00 Gyr) and Terzan 8 (13.50 Gyr) are old.
Furthermore, the five globular cluster companions to the Fornax dwarf spheroidal galaxy are neither unusually large nor of significantly below-average luminosity (van den Bergh 1994). It is noted in passing that NGC 6791, which with an age of $\sim$8 Gyr (Platais 2011) is the oldest known Galactic open cluste, has a luminosity $M_v$ = -4.6 that is comparable to that of the youngest globulars listed in Table 2.
However, its metallicity of [Fe/H] = +0.3 is an order of magnitude greater than that of the youngest Galactic globulars. This observation suggests (but does not prove) that there might not be a smooth transition from objects that we call Galactic open clusters to those that we regard as aglactic globular clusters. It is of some interest to note that NGC 6522, which Barbuy et al. (2009) claim to be the oldest known Galactic globular cluster, is situated in the dense nuclear bulge of the Galaxy at $R_{gc}$ = 0.6 kpc, whereas the youngest known globular clusters are located in the low-density outer halo of the Galaxy. In other words there may be some evidence for the assumption that the Galactic globular cluster system formed inside out.

\subsection{Cluster metallicities}

Figure 3 shows a plot of the metallicity [Fe/H] as a function of luminoisity $M_v$ for all Glactic globular clusters for which this information is given in Table 1. The figure shows a complete lack of correlation between the luminosities of globular clusters and their metallicities. Taken at face value this result appears to suggest that the luminosity (mass) with which a globular cluster is formed is entirely independent of the metallicity of the molecular cloud from which it was assembled.
Alternatively, typical globular clusters might have suffered different amounts of mass loss after their formation. It is also found that cluster metallicity is independent of the cluster concentration index C.
This result is somewhat surprising because one might, perhaps, have expected the most compact clusters to be concentrated in the metal-rich bulge region of the Galaxy.

Figure 4 plots the half-light radii $R_h$  of globular clusters as a function of their metallicities. The figure shows that half-light radii are weakly correlated with metallicity in the sense that the most compact clusters have the highest metallicities. All clusters with $R_h >$ 10 pc are seen to have [Fe/H] $<$ -1.4.

Figure 5 shows a surprisingly weak correlation between the
metallicity of globular clusters and their Galactocentric distances.
None of the clusters that are presently within 1.6 kpc of the Galactic center are more metal poor than [Fe/H] = -1.5. On the other hand only two clusters with $R_{gc} >$ 16 kpc [Whiting 1 and Palomar 1] have [Fe/H]$>$  -1.0.

\subsection{Cluster ellipticity}

There appear to be no clear-cut correlations between the
ellipticities of clusters and other parameters that describe them such as: (1) Cluster half-light radii $R_h$, (2)  central concentrations C, (3) metalicities [Fe/H], (4) Galactocentric distances $R_{gc}$ or the ages of globulars. The scatter in these parameters appears to be larger for flattened clusters with $\epsilon >$ 0.17 than it is for rounder clusters. However, the number of highly flattened clusters
(n=8) is too small to establish this result with confidence. An interesting difference between the star clusters in the Galaxy and those in the Magellanic Clouds is that highly flattened objects such as the LMC globular cluster NGC 1835, the SMC globular NGC 121 and the LMC massive open cluster NGC 1978, appear to be rare among both open and globular clusters in the Galaxy. It is noted in passing that inspection of the prints of the Palomar Sky Survey shows that highly flattened objects also appear to be absent from the population of Galactic $\it{open}$ clusters.

\subsection{Cluster luminosity}

Figure 6 shows a plot of cluster luminosity versus Galactocentric
distance. It should be noted that the data in this plot are probably somewhat less complete for clusters fainter than $M_v$ = -6 than they are for more luminous objects, For clusters with $R_{gc} <$ 40 kpc, $M_v$ and $R_{gc}$ appear to be uncorrelated. This observation presents some problems for theories of cluster evolution (Gieles et al. 2011). Furthermore, as has been known for may years (e.g. van den Bergh  2000a , p. 64),  globular clusters with $R_{gc} >$ 40 kpc are of below-average luminosity. The glaring exception to this rule is, of course, the luminous (and very distant) cluster NGC 2419 (shown as a + sign in Figure 6). Furthermore, many of these distant low-luminosity globulars appear to have ages (Sarajedini
2008) that are few Gyr younger than those in the main body of the Galaxy. For clusters that are more luminous than $M_v$ = -6.0 there seems to be no correlation between cluster luminosity and age. It is of interest to note that the outer halo of M31 (Huxor et al. 2005, 2011), does appear to contain a number of large globular clusters that have luminosities which are intermediate between those of NGC 2419 and the faint outer Galactic halo clusters. These differences between the characteristics of the M31 and Galaxy globular clusters suggest (van den Bergh 2006) that the Milky Way and the Andromeda galaxy accumulated from their ancestral objects in significantly different ways. Perhaps M31 experienced more massive (or later) merger events than did the Milky Way.

\subsection{Cluster half-light radii}

Figure 7 shows a plot of cluster luminosity $M_v$ versus half-light radius $R_h$. The figure shows two features: (1) Globular clusters in the main body of the Galaxy with $R_{gc} <$ 10 kpc (shown as blue dots) exhibit no obvious correlation between cluster luminosity and radius.
On the other hand clusters at $R_{gc} >$ 10 kpc, which have been plotted as red dots, are seen to be predominantly of below-average luminosity. The cluster NGC 2419 (shown as a + sign) deviates from this rule by being both very large and very luminous.

\subsection{Central concentration of light}

Figure 8 shows a loose correlation between the central concentration
of globular clusters and their luminosity, in the sense that the most luminous globular clusters also tend to be the most centrally concentrated ones.  The clusters with collapsed cores, which were arbitrarily assigned C = 2.5, do not fall on this relationship.   Such collapsed core clusters turn out to be of average luminosity.

\subsection{Cluster age}

The relation between cluster age and metallicity [Fe/H] depends
on cluster luminosity. Low-luminosity clusters with $M_v >$ -6.0 exhibit a large dispersion in age which makes it impossible to see if there is a residual correlation between their ages and their metallicities.
However, for clusters that are more luminous than $M_v$ = -6.0 there is a clear trend of decreasing metallicity with increasing age. Clusters with [Fe/H] = -0.5 have typical ages of $\sim$12.5 Gyr, compared to ages of $\sim$13.0 Gyr for [Fe/H] = -1.8. There is no clear-cut correlation between cluster age and the central concentration parameter C. In a plot of cluster age versus Log $R_h$ individual globulars fall into two distinct regions. The young clusters with ages $<$ 11.5 Gyr (i.e. the objects listed in Table 2) mostly have large half-light radii. The older clusters with ages $>$ 11.5 Gyr exhibit no obvious correlation between size and age. It is, however, of interest to note that the two most metal-rich clusters in Table 2 are also the smallest objects in this table.  The data in Table 1 show that young clusters are systematically more distant from the Galactic nucleus than are older clusters with ages $>$ 11.5 Gyr.  A Kolmogorov-Smirnov test shows that there is only a a 0.3\% probability that the old clusters with the age $>$ 11.5 Gyr were drawn from the same radial distribution as the younger clusters with ages $\leqslant$ 11.5 Gyr.  By the same token a K-S test shows that there is only a 1.3\% probability that the old clusters (which tend to be compact) were drawn from the same half-light radius distribution as the young clusters in this table.  Finally there is only a 1\% probability that the faint young clusters in Table 1 were drawn from the same parent luminosity distribution as the old clusters with ages $>$ 11.5 Gyr.

\subsection{Oosterhoff type}

Oosterhoff (1939) showed that Galactic globular clusters may be divided into two apparently distinct types on the basis of the properties of their RR Lyrae variables. Information on the assignment of clusters to Oosterhoff type I and Oosterhoff type II is available from recent papers by Catalan (2009), Kunder et al. (2011) and Amigo et al. (2011). Inspection of Table 1 shows a strong correlation between Oosterhoff type and the ages of Galactic globular clusters assigned by Dotter et al. (2010). Nine out of ten Oo II clusters have ages $\geq$ 13.0 Gyr, while 11 out of 12 Oo I clusters have ages $<$ 13.0 Gyr. The only cluster that does not fit into the pattern in which Oo II clusters are older than Oo I clusters is NGC 7089, to which Dotter et al.
assign an age of 12.5 Gyr. Inspection of the distribution of clusters of types Oo I and Oo II in Figures 1, 2, 3, 4, 5, 6, 7 and 8 shows that the only parameter that appears to correlate with Oosterfoff type is the metallicity [Fe/H]. Clusters of type Oo I have [Fe/H] $>$ -1.6, whereas those of type Oo II have [Fe/H] $<$ -1.6. The only small discrepancy is provided by the Oo type II cluster Omega Centauri (NGC 5139) which has [Fe/H] = -1.53. Possibly this deviation is due to the fact that NGC 5139 appears to have had an unusual evolutionary history. As a result the Helium abundance in Omega Cen might have become systematically larger than it is in typical Galactic globular clusters.
In summary $R_{gc}$, $M_v$, ellipticity, concentration index, and half-light radius do not appear to affect the period distribution of RR Lyrae stars. It is interesting to note (Smith 1995, p.119) that there is some evidence for a division into Oosterhoff groups for the globular clusters in the Magellanic Clouds, although the distinction into groups may be less clear-cut in the Clouds than it is within the Milky Way system. Individual RR Lyrae variables in various Local Group dwarf galaxies do not exhibit the Oosterhoff dichotomy (Smith 1995, p. 130).
The reason for this is, of course, that RR Lyrae variables in any individual dwarf galaxy can have widely differing ages and metallicities, whereas the ages and metallicities of stars within a particular globular cluster are mostly expected to exhibit only a small range in these parameters.

\section{CONCLUSIONS}

An almost complete sample of data on Galactic globular clusters
has been used to search for systematic effects relating the Galactocentric distances, metallicities, luminosities, ellipticities, central concentrations, half-light radii, and ages of these objects.
The data suggest that the Galaxy, the LMC and the Fornax dwarf spheroidal galaxy were already distinct stellar systems at the time when they formed their globular clusters. Differences between the systematics of globular clusters in the Galaxy and M31 suggest that these two galaxies were assembled from their ancestral objects in significantly different ways. It is found that the view that the outer halo globular clusters differ from those in the inner halo, because they were formed in ancestral objects that were subsequently captured, is too simpleminded. The previous finding that $R_h~ \alpha$~ $R^{2/3}_{gc}$ is confirmed. As a consequence of this relation between cluster size and Galactocentric distance, most clusters with collapsed cores are located close to the Galactic center. The only two exceptions are Palomar 1 and Palomar 12 for which the high measured central concentration might be due to errors resulting from the small stellar population samples contained in these objects. The luminosities of globular clusters are found to be independent of their metallicities.
Taken at face value this result suggests that the metallicity of molecular clouds does not affect the masses of the clusters that formed within them. Surprisingly the central concentration index C for globular clusters also appears to be independent of metallicity. This result is suprising because one might have expected the compact clusters in the Galactic bulge to have above-average metallicity. Within the main body of the Milky Way $(R_{gc} <$ 10 kpc) the luminosities and the radii of globular clusters appear to be uncorrelated. Striking differences between the Galaxy and the Magellanic Clouds are that: (1) some of the luminous globular and open clusters in the LMC and in the SMC are quite highly flattened, whereas globular clusters in the Milky Way system mostly appear to be nearly circular in outline. Inspection of the prints of the Palomar Sky Survey shows that the same conclusion applies to Galactic open clusters. Furthermore (2) at a given galactocentric distance LMC globulars are, on average, significantly larger than are Galactic globular clusters.

It is a pleasure to thank Bill Harris for making a pre-publication version of his catalog available to me. I am also thankful to him for wise advice on the first draft of the present paper. Thanks are also due to Alan McConnachie for reading and commenting on an early draft of this paper. I am also indebted to Alan McConnachie and Peter Stetson for providing important references. Thanks are also due to Brenda Parrish and Jason Shrivell for technical support.  Finally I thank the referee for his kind and helpful report.

\begin{deluxetable}{lrrllllllll}
\tablecaption{Data on Galactic globular clusters}

\tablehead{\colhead{ID}      & \colhead{$R_{gc}$}   & \colhead{log$R_{gc}$}   & \colhead{[Fe/H]}   & \colhead{$M_{v}$}    & \colhead{$\epsilon$}  & \colhead{C}      & \colhead{$R_{h}$}   &\colhead{log {$R_h$}}   & \colhead{age} & \colhead{Type}}

\startdata

N 104 &   7.4 &  0.87 &  -0.72  &  -9.42 &  0.09 &  2.07  &  4.15  &   0.62  &    12.75   &           \\
N 288 &  12.0 & 1.08  &-1.32  &  -6.75 &  ...  &  0.99  &  5.77   & 0.76  &    12.50   &              \\
N 362 &   9.4 &  0.97 &  -1.26  &  -8.43 &  0.01 &  1.76  &  2.05    &  0.31  &    11.50   &   I       \\
Whi 1 &  34.5 & 1.54    & -0.70 & -2.46  &   ... &   0.55 &   1.93  &   0.29 &            &                 \\
N1261 &  18.1 & 1.26   & -1.27 & -7.80  &   0.07&   1.16 &   3.22 &    0.51 &     11.50  &  I         \\
Pal 1 &  17.2 & 1.24    & -0.65 & -2.52  &   0.22&   2.57 &   1.49   &  0.17 &            &                  \\
AM 1  & 124.6 & 2.10   & -1.70 & -4.73  &   ... &   1.36 &   14.71  &   1.17 &   11.10   &               \\
Eri   &  95.0 & 1.98    &-1.43 & -5.13  &   ... &   1.10 &   12.06   &  1.08 &   10.50   &                   \\
Pal 2 &  35.0 & 1.54     &-1.42 & -7.97  &   0.05&   1.53 &   3.96   &   0.60 &           &                    \\
N1851 &  16.6 & 1.22   &   -1.18 & -8.33  &   0.05&   1.86 &   1.80  &    0.26 &           & I\\
N1904 &  18.8 & 1.27   &   -1.60 & -7.86  &   0.01&   1.70 &   2.44  &    0.39 &           & II\\
N2298 &  15.8 & 1.20   &  -1.92 & -6.31  &   0.08&   1.38 &   3.08   &  0.49 &   13.00   &  \\
N2419 &  89.9 & 1.95  &    -2.15 & -9.42  &   0.03&   1.37 &   21.38    & 1.33 &   13.00   & II\\
Ko 2  &  41.9 & 1.62   &    ...  & -0.35  &   ... &   0.50 &   2.12   &   0.33 &           &  \\
Pyx   &  41.4 & 1.62  &    -1.20 & -5.73  &   ... &    ... &    ...   &   ...  &           &  \\
N2808 &  11.1 & 1.05   &   -1.14 & -9.39  &   0.12&   1.56 &   2.23  &    0.35 &           &  I\\
E 3   &   9.1 & 0.96    &  -0.83 & -4.12  &   ... &   0.75 &   4.95   &   0.69 &           &  \\
Pal 3 &  95.7 & 1.98   &   -1.63 & -5.69  &   ... &   0.99 &   17.49   & 1.24 &   11.30   &   \\
N3201 &   8.8 & 0.94  &    -1.59 & -7.45  &   0.12&   1.29 &   4.42  &   0.65 &   12.00   &  I\\
Pal 4 & 111.2 & 2.05   &  -1.41 & -3.11  &   ... &   0.93 &   16.13   &  1.21 &   10.90   &   \\
Ko 1  &  49.3 & 1.69   &    ...  & -4.25  &   ... &   0.50 &   3.65   &   0.56 &           &   \\
N4147 &  21.4 & 1.33  &    -1.80 & -6.17  &   0.08&   1.83 &   2.69  &   0.43 &   12.75   &   \\
N4372 &   7.1 & 0.85    & -2.17 & -7.79  &   0.15&   1.30 &   6.60    &  0.82 &           &    \\
Ru 106&  18.5 & 1.27   &   -1.68 & -6.35  &   ... &   0.70 &   6.48   &   0.81 &           &    \\
N4590 &  10.2 & 1.01   &   -2.23 & -7.37  &   0.05&   1.41 &   4.52  &    0.66 &   13.00   & II  \\
N4833 &   7.0 & 0.85   &   -1.85 & -8.17  &   0.07&   1.25 &   4.63   &  0.67 &   13.00   &  II\\
N5024 &  18 4 &1.26   &    -2.10 &-8.71   &   0.01&   1.72 &   6.82  &    0.83 &   13.25   &   II\\ 
N5824 &  25.9 & 1.41  &    -1.91 & -8.85  &   0.03&   1.98 &   4.20    &  0.62 &           &  II\\
Pal 5 &  18.6 & 1.27     & -1.41 & -5.17  &   ... &   0.52 &   18.42    & 1.27 &           &   \\
N5897 &   7.4 & 0.87   &  -1.90 & -7.23  &   0.08&   0.86 &   7.49    & 0.87 &           &   \\
N5904 &   6.2 & 0.79   &   -1.29 & -8.81  &   0.14&   1.73 &   3.86  &   0.59  &   12.25   &   I\\
N5927 &   4.6 & 0.66   &   -0.49 & -7.81  &   0.04&   1.60 &   2.46  &    0.39 &   12.25   &    \\
N5946 &   5.8 & 0.76   &   -1.29 & -7.18  &   0.16&   2.50 &   2.74  v   0.44 &           &    \\
BH 176&  12.9 & 1.11  &     0.00 & -4.06  &   ... &   0.85 &   4.95    &  0.69 &           &    \\
N5986 &   4.8 & 0.68   &   -1.59 & -8.44  &   0.06&   1.23 &   2.96   &   0.47 &   13.25   &  II\\
Lyng 7&   4.3 & 0.63    &  -1.01 & -6.60  &   ... &   0.95 &   2.79     & 0.45 &   12.50   &    \\
Pal 14&  71.6 & 1.85   &   -1.62 & -4.80  &   ... &   0.80 &   27.15   &  1.43 &   10.50   &    \\
N6093 &   3.8 & 0.58   &   -1.75 & -8.23  &   0.00&   1.68 &   1.77  &    0.25 &   13.50   &     \\
N6121 &   5.9 & 0.77   &   -1.16 & -7.19  &   0.00&   1.65 &   2.77  &    0.44 &   12.50   &   I \\
N6101 & 2.5  0&.40    &    -0.74 & -4.82  &   ... &   0.70 &   2.98    &  0.47 &           &     \\
N6171 &   3.3 & 0.52  &    -1.02 & -7.12  &   0.02&   1.53 &   3.22    &  0.51 &   12.75   &   I \\
1636-2&   2.1 & 0.32  &    -1.50 & -4.02  &   ... &   1.00 &   1.21  &    0.08 &           &     \\
N6205 &   8.4 & 0.92   &   -1.53 & -8.55  &   0.11&   1.53 &   3.49    &  0.54 &   13.00   &     \\
N6229 &  29.8 & 1.47   &   -1.47 & -8.06  &   0.05&   1.50 &   3.19   &   0.50 &           &    I\\
N6218 &   4.5 & 0.65    &  -1.37 & -7.31  &   0.04&   1.34 &   2.47   &   0.39 &   13.25   &     \\
FRS173&   3.7 & 0.57   &    ...  & -6.45  &   ... &   0.56 &   0.97  &   -0.01 &           &      \\
N6235 &   4.2 & 0.62    &  -1.28 & -6.29  &   0.13&   1.53 &   3.35   &   0.53 &           &    \\
N6254 &   4.6 & 0.66   &   -1.56 & -7.48  &   0.00&   1.38 &   2.50   &   0.40 &   13.00   &    \\
N6256 &   3.0 & 0.48   &   -1.02 & -7.15  &   ... &   2.50 &   2.58   &   0.41 &           &    \\
Pal 15&  38.4 & 1.58    &  -2.07 & -5.52  &   ... &   0.60 &   14.43   &  1.16 &           &    \\
N6266 &   1.7 & 0.23   &   -1.18 & -9.18  &   0.01&   1.71 &   1.82  &    0.26 &           &    I\\
N6273 &   1.7 & 0.23    &  -1.74 & -9.13  &   0.27&   1.53 &   3.38   &   0.53 &           &      \\
N6284 &   7.5 &  0.88    & -1.26&   -7.96 &   0.03 &  2.50 &    2.94   &   0.47 &           &      \\
N6287 &   2.1 &  0.32    &  -2.10&    -7.36  &  0.13 &  1.38  &  2.02  &   0.31 &           &  \\
N6293 &   1.9 &  0.28   &   -1.99 &   -7.78  &  0.03 &  2.50  &  2.46  &   0.39  &          &   \\
N6304 &   2.3 &  0.36   &   -0.45 &  -7.30   &  0.02 &  1.80  &  2.44  &   0.39 &   12.75  &  \\
N6316 &   2.6 &  0.41    &  -0.45 &   -8.34  &  0.04 &  1.65  &   1.97 &    0.29 &          &    \\
N6341 &   9.6 &  0.98   &   -2.31 &   -8.21  &  0.10 &  1.68  &  2.46   &   0.39 &   13.25  &     II  \\
N6325 &   1.1 &  0.04   &   -1.25 &   -6.96  &  0.12 &  2.50  &  1.43   &   0.16  &         &     \\
N6333 &   1.7 &  0.23   &   -1.77 &   -7.95  &  0.04 &  1.25  &  2.21   &   0.34  &         &    II   \\
N6342 &   1.7 &  0.23   &   -0.55 &   -6.42  &  0.18 &  2.50  &  1.80  &    0.26  &         &      \\
N6356 &   7.5 &  0.88   &   -0.40 &  -8.51   &  0.03 &  1.59  &  3.56  &    0.55  &         &       \\
N6355 &   1.4 &  0.15   &   -1.37 &   -8.07  &  ...  &  2.50  &  2.36    &  0.37  &         &       \\
N6352 &   3.3 &  0.52   &   -0.64 &   -6.47  &  0.07 &  1.10  &  3.34   &   0.52  &   13.00 &      \\
I1257 &  17.9 &  1.25    &  -1.70 &   -6.15  &  ...  &  1.55  &  10.18   &  1.01  &         &             \\
Ter 2 &   0.8 & -0.10    &  -0.69 &   -5.88  &  ...  &  2.50  &  3.32    &  0.52  &         &          \\      
N6366 &   5.0 &  0.70   &   -0.59 &  -5.74   &  ...  &  0.74  &   2.92  &   0.47  &  12.00  &         \\
Ter 4 &   1.0 &  0.00    &  -1.41 &  -4.48   &  ...  &  0.90  &  3.87   &   0.59  &         &        \\
HP 1  &   0.5 & -0.30   &   -1.00 &   -6.46  &  ...  &  2.50  &  7.39  v   0.87  &         &        \\
N6362 &   5.1 &  0.71 &     -0.99 &   -6.95  &  0.07 &  1.09  &  4.03    &  0.61  &  12.50  &      I\\
Lil 1 &   0.8 & -0.10     & -0.33 &   -7.32  &  ...  &  2.30  &  ...   &    ...   &         &             \\
N6380 &   3.3 &  0.52 &     -0.75 &   -7.50  &  ...  &  1.55  &  2.35   &   0.37  &         &          \\
Ter 1 &   1.3 &  0.11   &  -1.03 &   -4.41  &  ...  &  2.50  &  7.44   &  0.87  &         &               \\
Ton 2 &   1.4 &  0.15   &   -0.70 &   -6.17  &  ...  &  1.30  &  3.10   &   0.49  &         &          \\
N6388 &   3.1 &  0.49   &   -0.55 &   -9.41  &  0.01 &  1.75  &  1.50   &   0.18  &         &       \\
N6402 &   4.0 &  0.60  &    -1.28 &   -9.10  &  0.11 &  0.99  &  3.52   &   0.55  &         &       I\\
N6401 &   2.7 &  0.43  &    -1.02 &   -7.90  &  0.15 &  1.69  &  5.89   &   0.77  &         &          \\
N6397 &   6.0 &  0.78  &    -2.02 &   -6.64  &  0.07 &  2.50  &  1.94   &   0.29  &  13.50  &          \\ 
Pal 6 &   2.2 &  0.34   &   -0.91 &   -6.79  &  ...  &  1.10  &  2.02    &  0.31  &         &          \\
N6426 &  14.4 &  1.16  &    -2.15 &   -6.67  &  0.15 &  1.70  &  5.51  &    0.74   &        &        II\\
Djo 1 &   5.7 &  0.76     & -1.51 &   -6.98  &  ...  &  1.50  &   ...   &   ...   &          &      \\
Ter 5 &   1.2 &  0.08    &  -0.23 &   -7.42  &  ...  &  1.62  &  1.45  &   0.16  &           &       \\
N6440 &   1.3 &  0.11 &     -0.36 &   -8.75  &  0.01 &  1.62   &  1.19  &   0.08  &           &       \\
N6441 &   3.9 &  0.59  &    -0.46 &   -9.63  &  0.02 &  1.74   &  1.92  &   0.28  &           &        \\
Ter 6 &   1.3 &  0.11     & -0.56 &   -7.59  &   ... &  2.50  &   0.87  &  -0.06  &          &        \\
N6453 &   3.7 &  0.57  &    -1.50 &   -7.22  &  0.09 &  2.50    & 1.48  &   0.07  &           &       \\
UKS 1 &   0.7 & -0.15  &    -0.64 &   -6.91  &   ... &   2.10   &  ...  &    ...  &          &        \\
N6496 &   4.2 &  0.62 &     -0.46 &   -7.20  &  0.16 &  0.70  &   3.35  &   0.53  &   12.00  &      \\
Ter 9 &   1.1 &  0.04    &  -1.05 &   -3.71  &  ...  &  2.50  &  1.61  &   0.21  &          &       \\
Djo 2 &   1.8 &  0.26    &  -0.65 &   -7.00  &  ...  &  1.50  &   ...   &   ...   &          &       \\
N6517 &   4.2 &  0.62  &    -1.23 &   -8.25  &  0.06 &  1.82   &  1.54  &   0.19  &          &       \\
Ter 10&   2.3 &  0.36   &  -1.00  &  -6.35   &  ...  &  0.75   &  2.62  &   0.42  &          &       \\
N6522 &   0.6 & -0.22  &   -1.34  &  -7.65   &  0.06 &  2.50  &   2.24  &   0.35  &          &       \\
N6535 &   3.9 &  0.59  &   -1.79  &  -4.75   &  0.08 &  1.33  &   1.68  &   0.23  &  13.25   &       \\
N6528 &   0.6 & -0.22  &   -0.11  &  -6.57   &  0.11 &  1.50  &   0.87  &  -0.06  &          &       \\
N6539 &   3.0 &  0.48 &    -0.63  &  -8.29   &  0.08 &  1.74   &  3.86  &   0.59  &          &       \\
N6540 &   2.8 &  0.45 &    -1.35  &  -6.35   &  ...  &  2.50    & ...   &   ...   &          &       \\
N6544 &   5.1 &  0.71 &    -1.40  &  -6.94   &  0.22 &  1.63  &   1.06  &   0.03  &          &       \\
N6541 &   2.1 &  0.32  &   -1.81  &  -8.52   &  0.12 &  1.86   &  2.31  &   0.36  &  13.25   &       \\
2MS 01&   4.5 &  0.65 &     ...   &  -6.11   &  ...  &  0.85   &  1.73  &   0.24  &          &        \\
ESO 06&  14.0 &  1.15 &    -1.80  &  -4.87   &  ...  &  0.90  &   6.54  &   0.82  &          &       \\
N6553 &   2.2 &  0.34   & -0.18  &  -7.77   &  0.17 &   1.16  &  1.80  &   0.26  &          &       \\
2MS 02&   3.2 &  0.51   &  -1.08  &  -4.86   &  ...  &   0.95  &  0.78  &  -0.11  &          &       \\
N6558 &   1.0 &  0.00  &   -1.32  &  -6.44   &  ...  &   2.50  &  4.63  &   0.67  &          &     I\\
I1276 &   3.7 &  0.57  &   -0.75  & -6.67    &  ...  &   1.33  &  3.74  &   0.57  &          &       \\
Ter 12&   3.4 &  0.53   &  -0.50  & -4.14    &  ...  &   0.57   & 1.05  &   0.02  &          &       \\
N6569 &   3.1 &  0.49  &   -0.76  &  -8.28   &  0.00 &   1.31  &  2.54  &   0.40  &          &       \\
BH 261&   1.7 &  0.23  &   -1.30  &  -4.19   &  0.03 &   1.00  &  1.04  &   0.02  &          &       \\
GLI 02&   3.0 &  0.48   &  -0.33  &   ...    &   ... &   1.33  &  ...   &   ...   &          &                  \\
N6584 &   7.0 &  0.85   &  -1.50  &  -7.69   &   ... &   1.47 &   2.87  &   0.46  & 12.25    &    I  \\
N6624 &   1.2 &  0.08 &    -0.44  &  -7.49   &  0.06 &   2.50  &  1.88  &   0.27  & 13.00    &       \\
N6626 &   2.7 &  0.43 &    -1.32  &  -8.16   &  0.16 &   1.67   & 3.15  &   0.50  &          &     I  \\
N6638 &   2.2 &  0.34 &    -0.95  &  -7.12   &  0.01 &   1.33   & 1.39  &   0.14  &          &       \\
N6637 &   1.7 &  0.23  &   -0.64  &  -7.64   &  0.01 &   1.38  &  2.15  &   0.33  & 12.50    &       \\
N6642 &   1.7 &  0.23  &   -1.26  &  -6.66   &  0.03 &   1.99  &  1.72  &   0.24   &         &   I   \\
N6652 &   2.7 &  0.43  &   -0.81  &  -6.66   &  0.20 &   1.80   & 1.40  &   0.15   & 13.25   &       \\
N6656 &   4.9 &  0.69 &    -1.70  &  -8.50   &  0.14 &   1.38   & 3.13  &   0.50    &        &   II  \\
Pal 8 &   5.5 &  0.74  &   -0.37  &  -5.51   &  ...  &   1.53  &  2.16  &   0.33   &         &       \\
N6681 &   2.2 &  0.34 &    -1.62  &  -7.12   &  0.01 &   2.50  &  1.86  &   0.27   & 13.00   &        \\
GLI 01&   4.9 &  0.69    &  ...   &  -5.91   &  ...  &   1.37  &  0.79  &  -0.10   &         &       \\
N6712 &   3.5 &  0.54  &  -1.02  &  -7.50   &  0.11 &   1.05 &  2.67  &   0.43   &         &      I\\
N6715 &  18.9 &  1.28   &  -1.49  &  -9.98   &   0.06&   2.04 &   6.32  &   0.80   &         &       \\
N6717 &   2.4 &  0.38    & -1.26  &  -5.66   &  0.01 &   2.07  &  1.40  &   0.15   & 13.00   &       \\
N6723 &   2.6 &  0.41   &  -1.10  &  -7.83   &  0.00 &   1.11   & 3.87  &   0.59   & 12.75   &     I  \\
N6749 &   5.0 &  0.70  &   -1.60  &  -6.70   &   ... &   0.79  &  2.53  &   0.40   &         &         \\
N6752 &   5.2 &  0.72  &   -1.54  &  -7.73   &  0.04 &   2.50  &  2.22  &   0.35   & 12.50   &         \\
N6760 &   4.8 &  0.68  &   -0.40  &  -7.84   &  0.04 &   1.65  &  2.73  &   0.44   &         &          \\
N6779 &   9.2 &  0.96   &  -1.98  &  -7.41   &  0.03 &   1.38  &  3.01  &   0.48   & 13.50   &          \\
Ter 7 &  15.6 &  1.19    & -0.32  &  -5.01   &  ...  &   0.93  &  5.11  &   0.71   & 8.00    &           \\
Pal 10&   6.4 &  0.81  &   -0.10  &  -5.79   &  ...  &   0.58   & 1.70  &   0.23   &         &          \\
Arp 2 &  21.4 &  1.33 &    -1.75  &  -5.29   &  ...  &   0.88  &  14.73 &   1.17   &  13.00  &          \\
N6809 &   3.9 &  0.59 &    -1.94  &  -7.57   &  0.02 &   0.93  &  4.45  &   0.65   & 13.50   &         \\
Ter 8 &  19.4 &  1.29  &   -2.16  &  -5.07   &  ...  &   0.60  &  7.27  &   0.86   & 13.50   &        \\
Pal 11&   8.2 &  0.91  &   -0.40  &  -6.92   &  ...  &   0.57  &  5.69  &   0.76   &         &        \\
N6838 &   6.7 &  0.83   &  -0.78  &  -5.61   &  0.00 &   1.15   & 1.94  &   0.29   & 12.50   &          \\
N6864 &  14.7 &  1.17  &   -1.29  &  -8.57   &  0.07 &   1.80  &  2.80  &   0.45   &         &          \\
N6934 &  12.8 &  1.11  &   -1.47  &  -7.45   &  0.01 &   1.53  &  3.13  &   0.50   & 12.00   &       I\\
N6981 &  12.9 &  1.11  &   -1.42  &  -7.04   &  0.02 &   1.21  &  4.60  &   0.66   & 12.75   &       I\\
N7006 &  38.5 &  1.59  &   -1.52  &  -7.67   &  0.01 &   1.41  &  5.27  &   0.72   &         &       I\\  
N7078 &  10.4 &  1.02  &   -2.37  &  -9.19   &  0.05 &   2.29  &  3.03  &   0.48   &  13.25  &       II\\
N7089 &  10.4 &  1.02  &   -1.65  &  -9.03   &  0.11 &   1.59  &  3.55  &   0.55   &  12.50  &      II\\
N7099 &   7.1 &  0.85   & -2.27  &  -7.45   &  0.01 &   2.50   & 2.43  &   0.39   & 13.25   &        \\
Pal 12&  15.8 &  1.20    & -0.85  &  -4.47   &  ...  &   2.98    &9.51  &   0.98   &  9.50   &        \\
Pal 13&  26.9 &  1.43    & -1.88  &  -3.76   &  ...  &   0.66  &  2.72  &   0.43   &         &        \\
N7492 &  25.3 &  1.40   &  -1.78  &  -5.81   &  0.24 &  0.72 &   8.80   &   0.94   &         &        \\

\enddata
\end{deluxetable}

\begin{deluxetable}{lrrrr}
\tablecaption{Clusters younger than 11.5 Gyr}
\tablehead{\colhead{Name} &\colhead{$R_{gc}$(kpc)} &\colhead{$R_h$(pc)} &\colhead{$M_v$} &\colhead{[Fe/H]}}

\startdata
AM 1     &  125  &   15   & -4.7  &  -1.70\\
Eridanus &  95   &   12   & -5.1  &  -1.43\\
Pal 3    &  96   &   17   & -5.7  &  -1.63\\
Pal 4    & 111   &   16   & -3.1  &  -1.41\\
Pal 14   &  72   &   27   & -4.8  &  -1.62\\
Terzan 7 &  16   &   5    & -5.0  &  -0.32\\
Pal 12   &  16   &   10   & -4.5  &  -0.85\\
\enddata
\end{deluxetable}

\newpage

\begin{figure}
\plotone{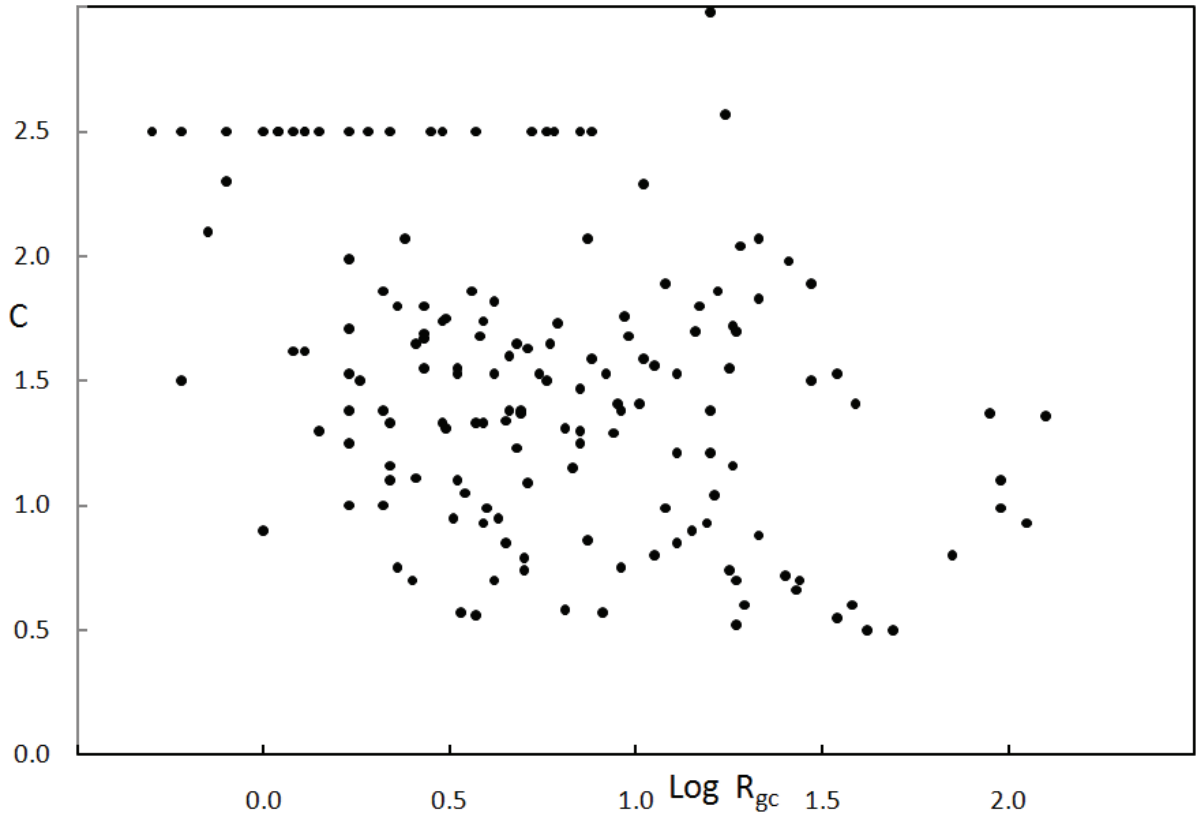}
\caption{Concentration index C as a function of Galactocentric distance. The figure shows that (1) Clusters with collapsed cores are most strongly concentrated to the Galactic nucleus, and (2) the clusters with Galactocentric radii in excess of 30 kpc all have low central concentrations of light.}
\end{figure}

\begin{figure}
\plotone{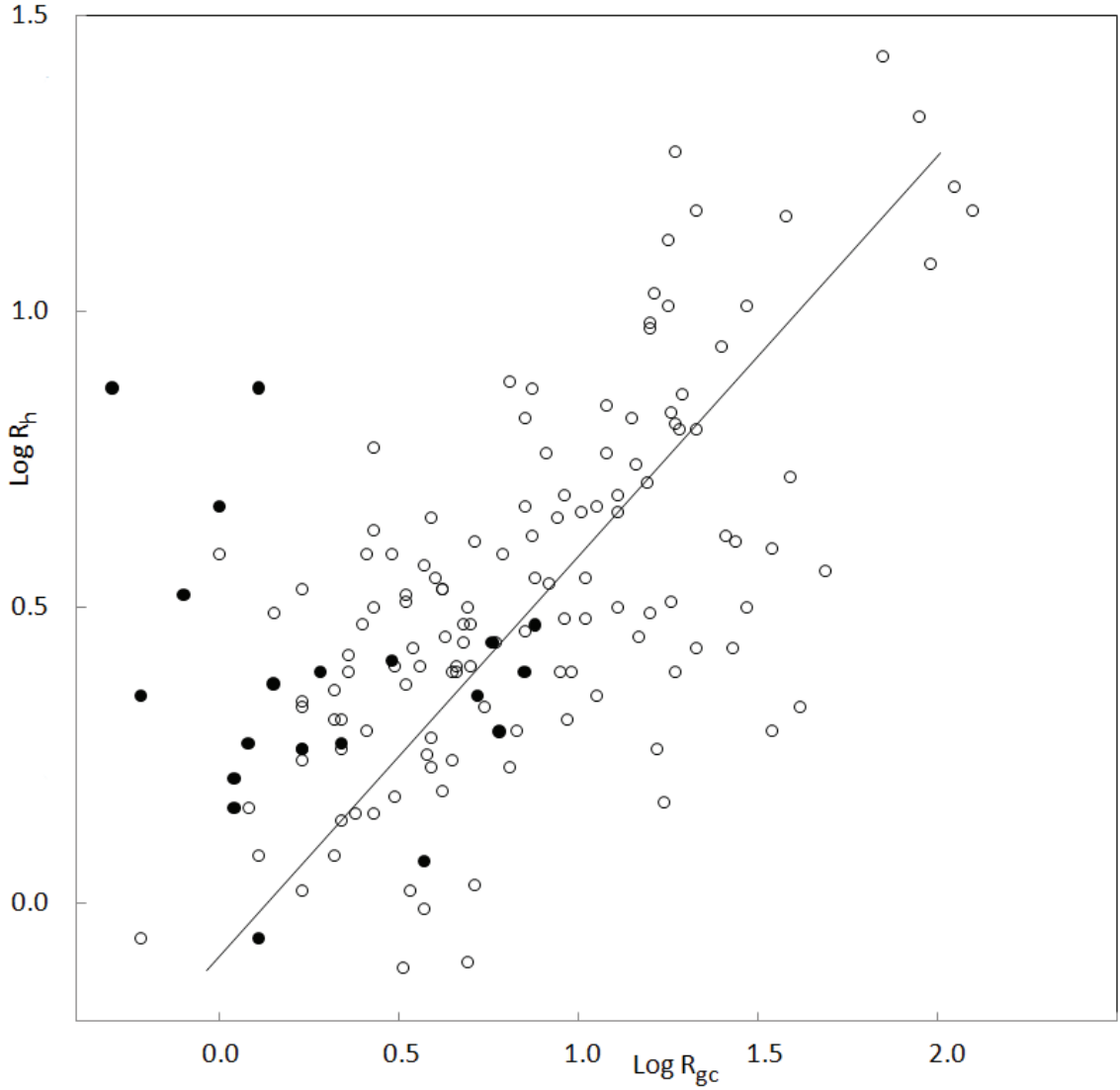}
\caption{Half-light radius versus Galactocentric distance. The figure shows that the half-light radii of globular clusters tend to increase with increasing Galactocentric distance. Clusters with collapsed cores (filled dots) are seen to favor small Galactocentric distances.}
\end{figure}

\begin{figure}
\plotone{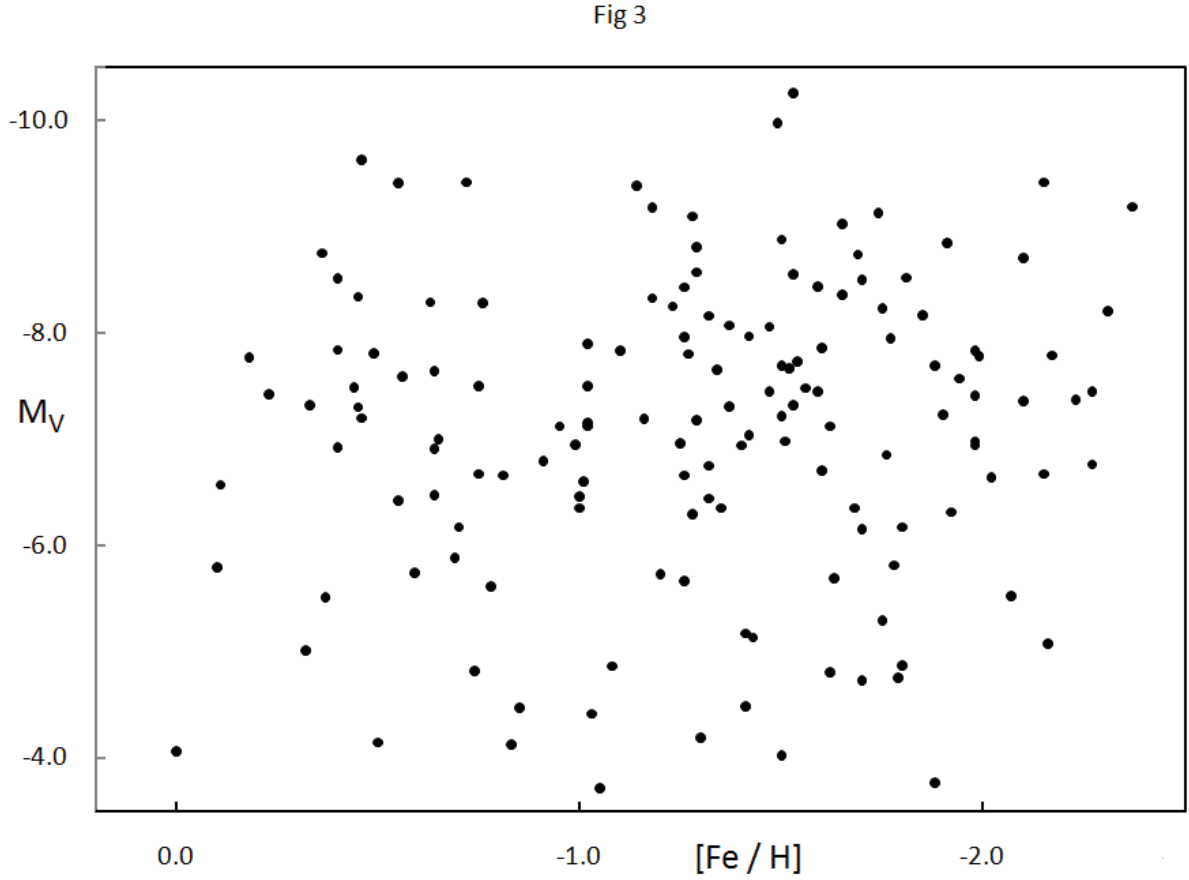}
\caption{This plot appears to show that the luminosities, and hence presumably the masses, with which globular clusters were formed are entirely independent of the metallicity of the gas from which they were formed.}
\end{figure}

\begin{figure}
\plotone{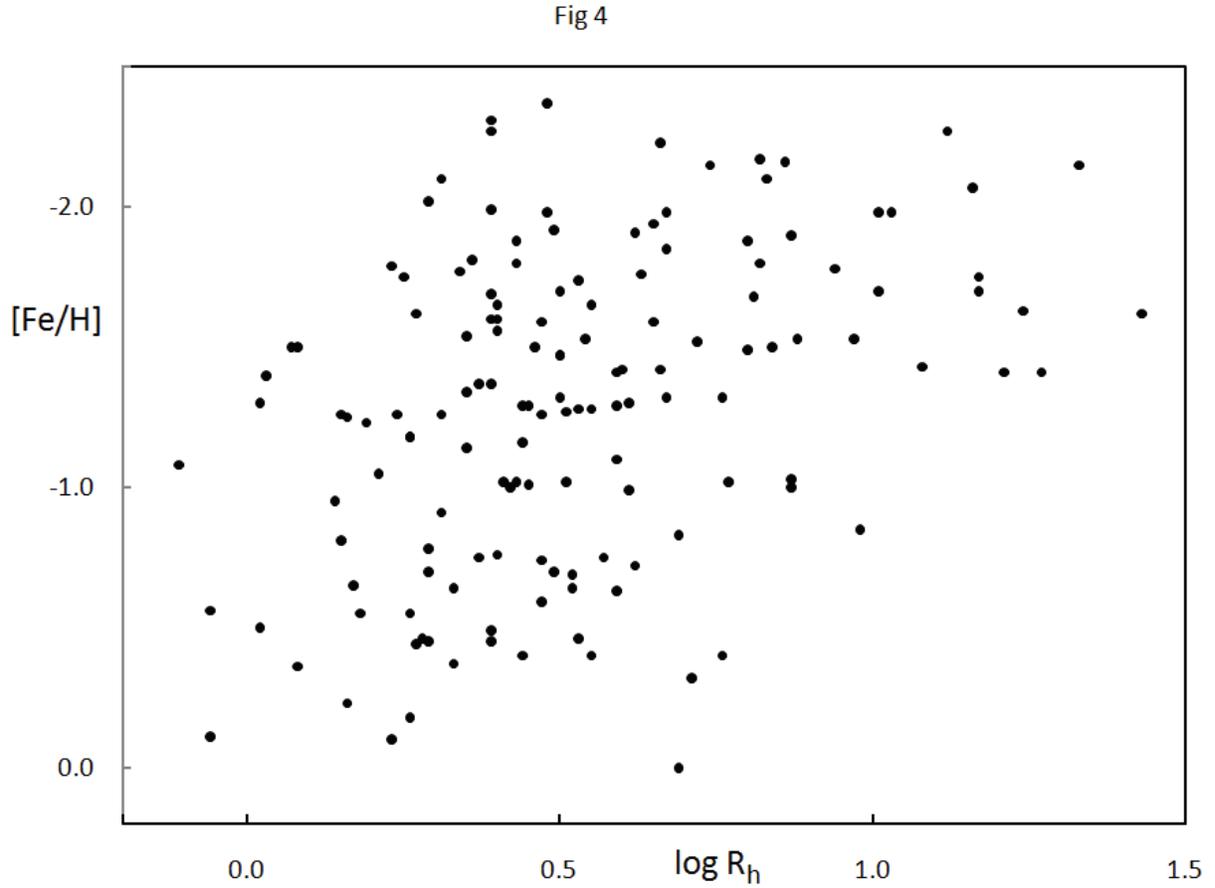}
\caption{Relation between the half-light radii of globular clusters and their metallicities. The figure shows a weak correlation in the sense that the most compact Galactic globular clusters have the highest metallicities. All clusters with $R_h >$ 10 pc are seen to have [Fe/H]$<$  -1.4.}
\end{figure}

\begin{figure}
\plotone{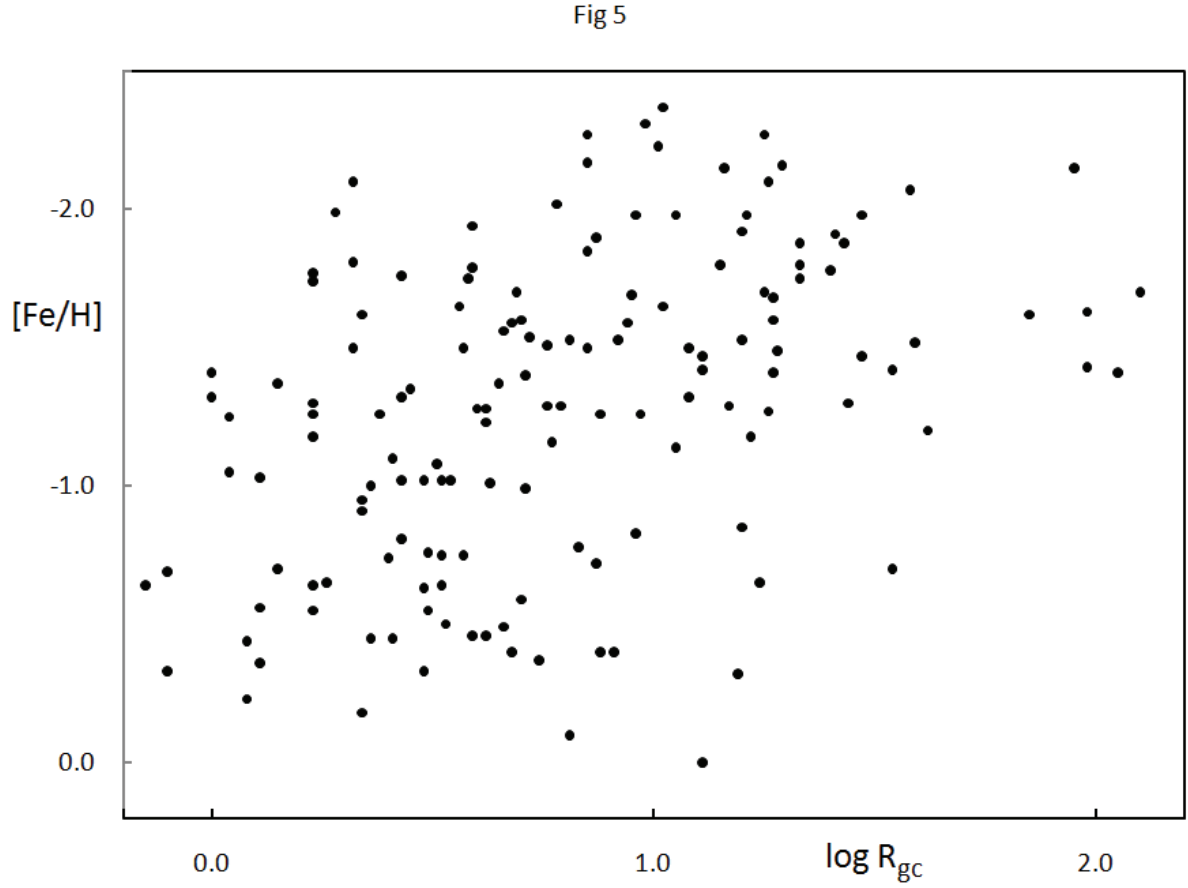}
\caption{Relation between Galactocentric distance $R_{gc}$ and metallicity [Fe/H] of Galactic globular clusters. Surprisingly only a relatively weak correlation is seen between the metallicity of globular clusters and their distance from the center of the Galaxy.}
\end{figure}

\begin{figure}
\plotone{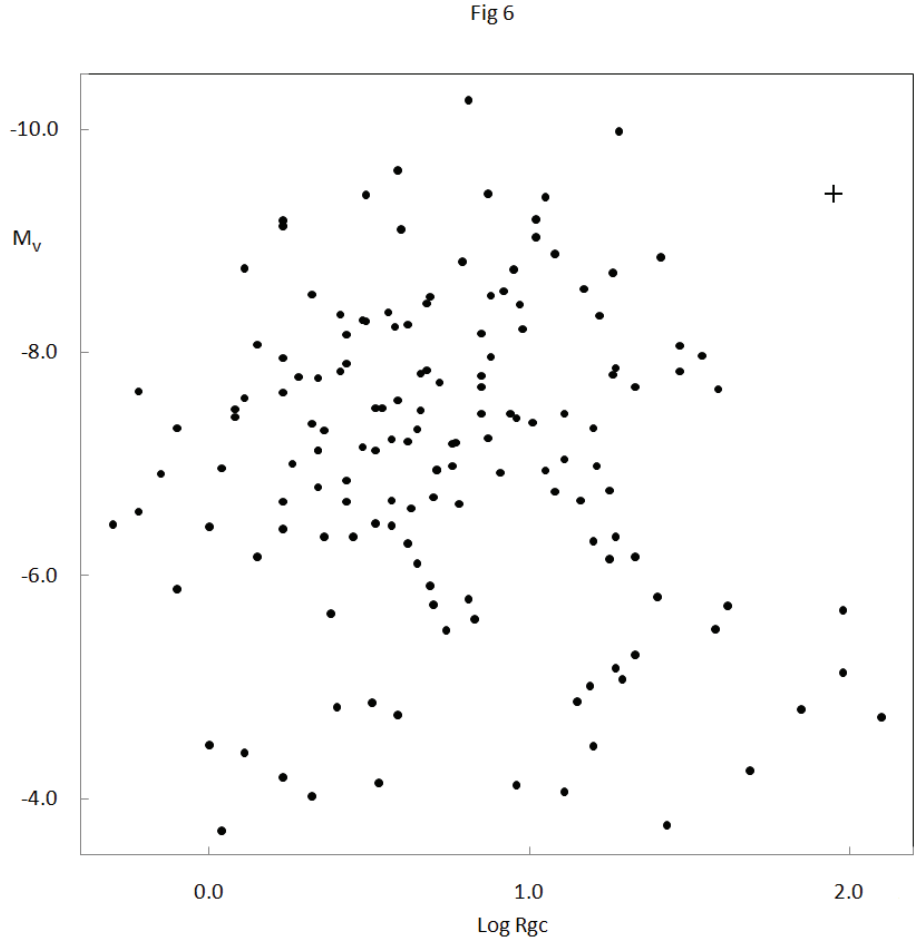}
\caption{Cluster luminosity versus Galactocentric distance.
With the single exception of NGC 2419 (shown as a + sign) clusters with $R_{gc} >$ 40 kpc are seen to be of below-average luminosity.}
\end{figure}

\begin{figure}
\plotone{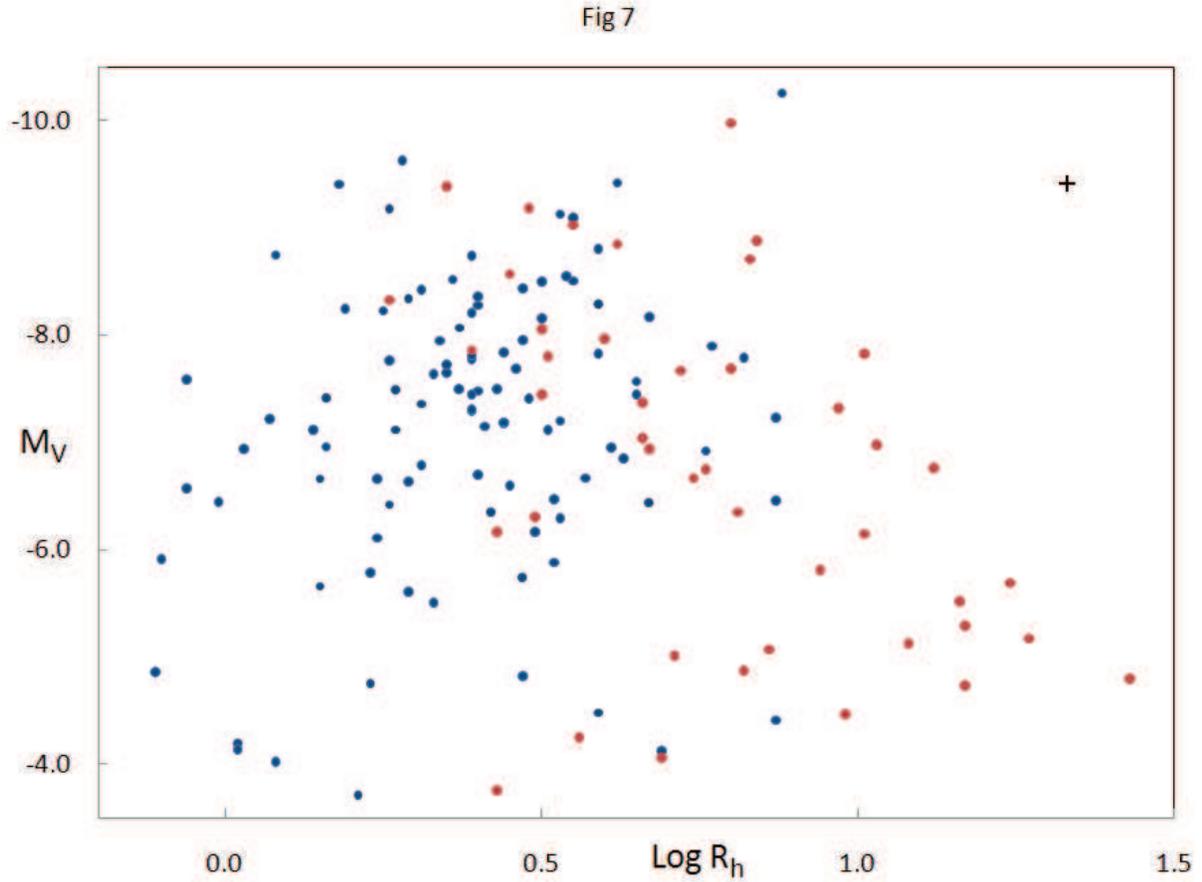}
\caption{Cluster luminosity $M_v$ versus half-light radius  $R_h$  for galactic globular clusters. Globular clusters in the main body of the Galaxy $(R_{gc} <$ 10 kpc), which are shown as blue dots, show no obvious correlation between luminosity and half-light radius.
On the other hand globulars with $R_{gc} >$ 10 kpc are mostly of below-average luminosity. An exception is the large luminous cluster NGC 2419 (which is plotted as a + sign).}
\end{figure}

\begin{figure}
\epsscale{0.6}\plotone{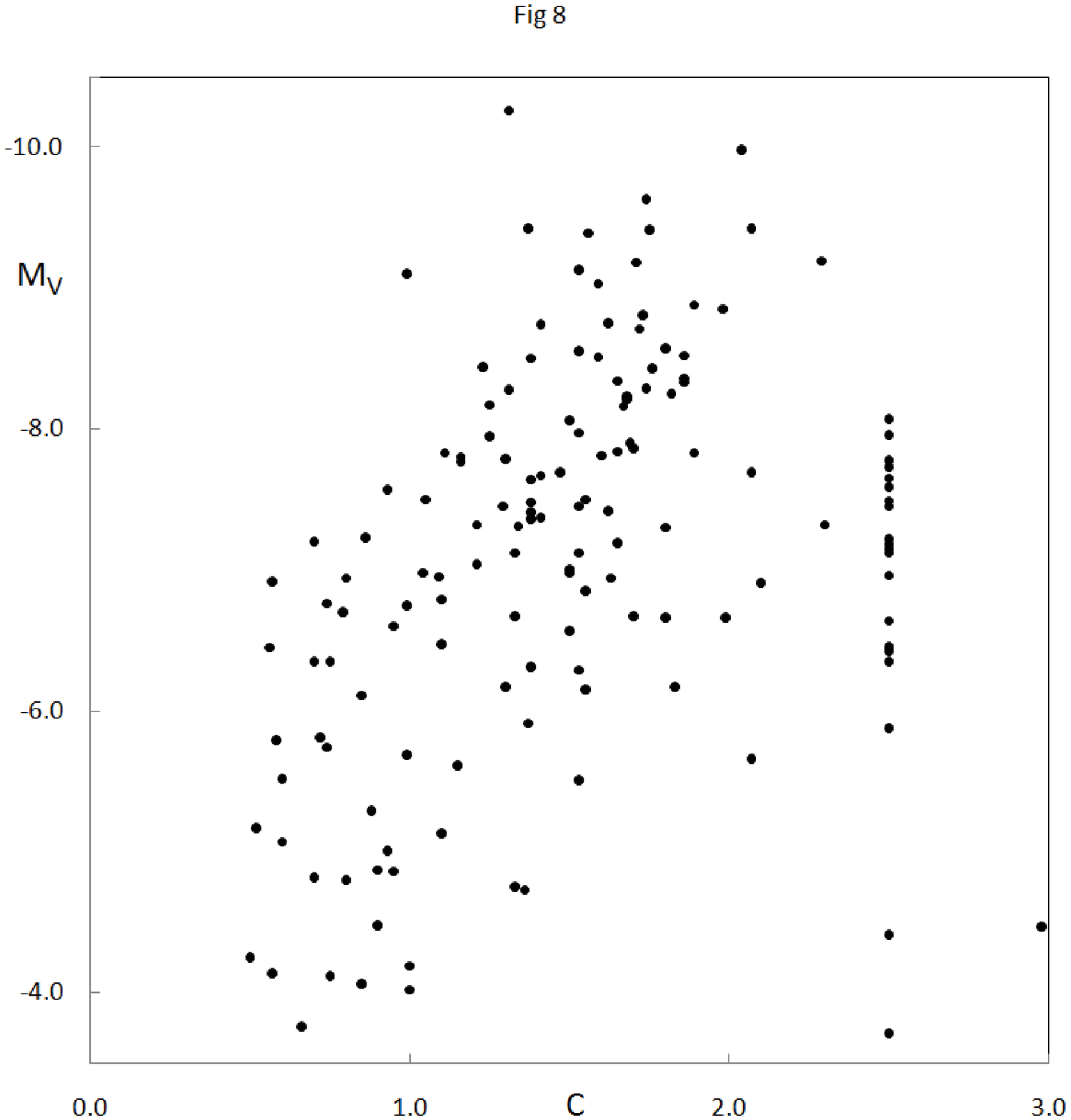}
\caption{Cluster luminosity and concentration are seen to be loosely correlated in the sense that the most luminous clusters also tend to be the ones with the highest central concentration of light. Surprisingly clusters with collapsed cores do not follow this correlation since they are only of average luminosity.}
\end{figure}


\begin{references}
\reference{}Amigo, P., Catelan, M., Stetson, P. B., Smith, H. A., Cacciari, C., \& Zoccali, M. 2011, Carnegie Observatories Astrophysics Series, Vol.5 (Pasadena: Carnegie Observatories = arXiv: 1105.0896)

\reference{}Arp, H. C. 1965, in Galactic Structure, Eds. A. Blaauw and .  M. Schmidt (Chicago: Univ. of Chicago Press),p. 401 
\reference{}Bailin, J. \&  Harris, W. E. 2009, ApJ, 695, 1082 


\reference{}Barbuy, B., Zoccali, M., Ortolani, S., Hill, V., Minniti, D., Bica, E.Renzini, A.\& G\'{o}mez, A. 2009, A\&A, 507, 405

\reference{}Bekki, K. \&  Freeman, K. C. 2003, MNRAS, 346, L11 

\reference{}Bellazzini, M. 1998, New Astronomy, 3, 219 

\reference{}Catelan, M. 2009, Astroph. \& Space Sci., 320, 261 

\reference{}Correnti, M., Bellazzini, M., Dalessandro, E., Mucciarelli, A . Monaco, L. \& Catelan, M. 2011, MNRAS (in press = arXiv:1105.2001 

\reference{}Djorgovski, S. 1995, ApJ, 438, L29 

\reference{}Djorgovski, S \& Meylan, G. 1993 in Structure and Dynamics of  Globular Clusters = ASP Conference Series No.50, Eds. 
S. G. Djorgovsski \& G. Meylan, San Francisco: ASP,  p. 325 

\reference{} Dotter, A. et al. 2010, ApJ, 708, 698 

\reference{} Fadely, R., Willman, B., Geha, M., Walsh, S., Mu$\tilde{n}$oz, R. R., Jerjen, H., Vargas, L. C. \& Da Costa, G. S., 2011,  AJ (in press = zrXiv:1107.3151)

\reference{}Ferraro, F. 2011 (in press = arXiv:1012.3616)

\reference{} Gieles, M., Heggie, D. C. \& Zhao, H-S. MNRAS (in press = arXiv:1101.1821 

\reference{}Harris, W. E. 1996, AJ, 112, 1487

\reference{}Huxor, A. P., Tanvir, N. R., Irwin, M. J., Ibata, R., Collett, J.L. Ferguson, A. M. N., Bridges, T. \& Lewis, G. F. 2005, MNRAS, 360, .  1007 

\reference{}Huxor, A. P. et al. 2011, MNRAS (in press = arXiv:1102.0403) 

\reference{}Ibata, R. A., Gilmore, G.\&  Irion, M. J. 1994, Nature, 370,194 

King, I. R. 1966, AJ, 71, 64 Koposov, S. et al. 2007, ApJ, 669, 337 

\reference{}Kukarkin, B. V. 1974, Globular Star Clusters (Moscow: Sternberg State Astron. Inst.)

\reference{}Kunder, A., Stetson, P. B., Catelan, M, Amigo, P, \&  De Propris, R. 2011, Carnegie Observatory Astrophysics Series, .  vol. 5: RR Lyrae Stars, Metal-Poor Stars and the Galaxy, .  ed. A. McWilliam (Pasadena: Carnegie Observatories = arXix .  1105.0008) 

Lightman, A. P. \&  Shapiro, S. L. 1978, Rev. Mod. Phys., 50, 437 

\reference{}Minniti, D. et al. 2011, A\&A, in press = arXiv: 1012.2450 

Murphy, B. W., Cohn, H. N. \&  Hut, P. 1990, MNRAS, 245, 355 

\reference{}Olszewski E. W., Saha, A., Knezek, P., Subramanian, A., de Boer, . T \& Seitzer, P. 2009, AJ, 138, 1570 

\reference{}Ortolani, S., Barbuy, B., Momany, Y., Saviane, I., Bica, E.Jilkova, L. Salerno, G. M. \& Jungwiert, B. 2011, arXiv: 1106.2725 

\reference{}Ortolani, S., Bica, E. \&  Barbuy, B. 1997, MNRAS, 284, 692 

\reference{}Pasquato, M. \&  Bertin, G. 2010, A\&A, 512, A35 

\reference{}Platais, I., Cudworth, K. M., Kozhurina-Platais, V. McLaughlin, D. E., Meibom, S. \&  Veillet, C. 2011, ApJ, 733,L1

\reference{}Sarajedini, A. 2008, in The Ages of Stars = IAU Symposium 258 E. E. Mamajek, D. R. Sonderblom \&  R. F. G. Wyse Eds. p.221

\reference{}Sawyer Hogg, H. 1959, Handbuch der Physik, 53, 129 

\reference{}Shapley, H. 1918, ApJ, 48, 154 Smith, H. A. 1995, RR Lyrae Stars (Cambridge: Cambridge University .  Press).

\reference{}Spitzer, L. \&  Thuan, T. X. 1972, ApJ, 175, 31 

\reference{}Surdin, V. G. 1994, Astr. Letters, 20, 15 

\reference{}van den Bergh, S. 1994, AJ, 108, 2145 

\reference{}van den Bergh, S. 2000a, The Galaxies of the Local Group (Cambridge: Cambridge University Press)

\reference{}van den Bergh, S. 2000b, AJ, 530, 777

\reference{}van den Bergh, S. 2006 in The Local Group as an Astrophysical    .
Laboratory, Eds. M. Livio and T. M. Brown (Cambridge: Cambridge University Press) p.1

\reference{}Webbink, R. F. 1985 in Dynamics of Star Clusters = IAU Symposium .  No. 113, Eds. J. Goodman \& P. Hut, (Dordrecht:  Reidel),p. 541
\end{references}
\end{document}